\documentclass[pra,twocolumn,superscriptaddress,aps]{revtex4}
\usepackage{amsfonts}
\usepackage{amsmath}
\usepackage{amssymb}
\usepackage{graphicx,color}
\usepackage{dcolumn}
\usepackage{times}
\usepackage{amsmath}

\DeclareMathOperator{\sgn}{sgn}
\DeclareMathOperator{\sech}{sech}

\begin{document}

\title{Dark Soliton Scattering in Symmetric and Asymmetric Double
Potential Barriers}

\author{F. Tsitoura}
\affiliation{Dynamics Group, Hamburg University of Technology, 
21073 Hamburg, Germany}

\author{Z. A. Anastassi}
\affiliation{Department of Mathematics, Statistics and Physics,
College of Arts and Sciences, Qatar University, 2713 Doha, Qatar}

\author{J. L. Marzuola}
\affiliation{Department of Mathematics, University of North Carolina,
Chapel Hill, NC 27599, USA}

\author{P. G. Kevrekidis}
\affiliation{Department of Mathematics and Statistics, University of Massachusetts,
Amherst, Massachusetts 01003-4515 USA}

\affiliation{Center for Nonlinear Studies and Theoretical Division, Los Alamos
National Laboratory, Los Alamos, NM 87544}

\author{D. J. Frantzeskakis}
\affiliation{Department of Physics, 
National and Kapodistrian University of Athens,
Panepistimiopolis, Zografos, Athens 15784, Greece}

\begin{abstract}
  Motivated by the recent theoretical study of (bright) soliton diode
  effects in systems with multiple scatterers, as well as by
  experimental investigations of soliton-impurity interactions,  
we consider some prototypical case examples of interactions
  of dark solitons with a pair of scatterers. In a way fundamentally
  opposite to the case of 
  bright solitons 
  (but consonant to their ``anti-particle character''),
  we find that 
  dark solitons accelerate as they pass the first
  barrier and hence cannot be trapped by a second equal-height barrier.
  A pair of unequal barriers may lead to reflection from the second
  one, however trapping in the inter-barrier region cannot occur. 
 We also give some examples of dynamical adjusting the barriers
  to trap the dark soliton in the inter-barrier region, yet we show
  that this can only occur over finite time horizons, with the
  dark soliton always escaping eventually, contrary again to what
  is potentially the case with bright solitons.
\end{abstract}

\maketitle

\section{Introduction}

One of the principal themes within 
studies of solitons in various physical contexts 
concerns their interactions with impurities,  
as well as with potential steps and barriers,  
within their domain of propagation~\cite{RMP}. In addition to the intrinsic
interest of such an area within nonlinear wave 
theory~\cite{delta_cauchy,holmer},
it is a subject broadly relevant, e.g., to solid state physics~\cite{kos1,kos2},
and nonlinear optics~\cite{newell,holmes}, and it has thus
been studied
both for bright~\cite{holmer2} and 
dark~\cite{vvk1} solitons.
In recent years, developments in the physics of atomic 
Bose-Einstein condensates (BECs)
~\cite{pita,book_fr,nl_fr,bec_20} have provided a fertile
playground where such studies can not only be theoretically
extended~\cite{gt,nn,nn2,greg,sch,pantofl,pantofl1,gardiner,martin,vas,fot},
but also experimentally explored. In fact, 
the combined presence of solitons and defects
in BECs
has been considered in experimentally relevant 
setups both for dark~\cite{eng,hul} (where the motion
in the presence of defects has been used to produce the solitons) and  
bright~\cite{dic,randy},
as well as for dark-bright~\cite{exp_eng}
solitons, in one- and multi-component BECs, respectively.

An intriguing variant of the problem that has been
explored recently, 
is that of inducing soliton trapping and transmission
from a pair of defects~\cite{Al}. 
A principal finding of the work of Ref.~\cite{Al}, where 
a bright soliton was scattered from the defects, was the following. 
As the soliton passed the
first defect, it lost some of its energy into radiation,
resulting in a reflection from the second one (even if
it was of equal size as the first), if the energy was
barely over the critical one needed to ``overcome'' the first defect.
This could be used in a two-fold way: 
considering a weakly asymmetric barrier pair, i.e., a slightly lower on the left
and a slightly higher on the right, it was possible to have a
bright soliton 
of the same energy get reflected 
from the pair of the two barriers when coming from the left,
but get transmitted when incident from the right. This
principal feature of this configuration was dubbed
in the work~\cite{Al} a ``soliton diode'' effect.
Similar diode features, but for linear wavepackets in
lattices with an asymmetric pair of nonlinear
elements, were also studied in Refs.~\cite{lepri,lepri1}.
However, an additional remarkable feature of the work of Ref.~\cite{Al}
was the possibility of trapping of a soliton within a region
of two equal barriers. Here, the loss of energy for a weakly
``supercritical'' barrier interaction results in the soliton
energy being ``subcritical'' with respect to the second barrier
and hence unable to overcome it. As a result, the 
soliton remains forever trapped in the region between the two defects.

Our aim in the present work is to study the scattering dynamics 
of a dark soliton by two potential barriers, and examine the possibility of 
soliton trapping, as per the spirit of Ref.~\cite{Al}. 
We find  
that dark solitons behave
in a fundamentally different way than bright ones: 
while bright solitons lose energy and decelerate, being more prone to trapping, 
dark ones accelerate, thus being amenable to escape dynamics. 
We find that, while a temporary form of trapping may be enforced by an asymmetric barrier
configuration, this is only short lived, because the dark soliton keeps 
accelerating with each collision, until it eventually
overcomes one of the two barriers. 

Our presentation of these results will be structured as follows.
In section II, we will provide a dynamical systems (theoretical)
analysis of the soliton-two barrier interaction. Then, in section III, we will
complement this analysis by means of direct numerical simulations.
Finally, in section IV, we will provide a summary of our results, as well as
number of directions for future study.
\\

\section{Theoretical Analysis: Particle Dynamics}

We start our analysis by presenting our model, originating from the
context of atomic BECs in the mean-field picture \cite{pita}. 
We consider a quasi one-dimensional (1D) setting whereby a BEC is oriented along
the $x$-direction and is confined in a strongly anisotropic
(quasi-1D) trap. In such a settting, the macroscopic wave function 
$u(x,t)$ satisfies the following 
dimensionless, 1D Gross-Pitaevskii equation~\cite{pita,book_fr,nl_fr,bec_20}:
\begin{eqnarray}
i u_t = -\frac{1}{2} u_{xx} 
+ s |u|^2  u + V(x) u, 
\label{nls}
\end{eqnarray}
where subscripts denote partial derivatives, 
$s=+1$ ($s=-1$) corresponds to respulsive (attractive) interatomic interactions, 
while $V(x)$ is the external potential.  
In our setting, the latter is assumed to consist of two Gaussian barriers, namely: 
\begin{eqnarray}
V(x)=\sum_{i=1}^2  \Bigg\lbrace V_i  \exp{\left[-\frac{1}{2}\left(\frac{x-l_i}{\sigma_i }\right)^2\right]}  \Bigg\rbrace, 
\label{Vg}
\end{eqnarray}
where $\sigma_i$ and $V_i$ set their widths and amplitudes, respectively, while 
$l_i$ denotes the position of their respective centers. Such a potential may be 
induced by a pair of far-detuned laser beams, featuring the most natural beam shape, namely  
the Gaussian profile \cite{book_fr}. Notice that the case $V_i>0$ ($V_i<0$) corresponds to 
blue- (red-) detuned laser beams, that repel (attract) the atoms in the condensate. 
It is also relevant to mention that in the limit $\sigma_i \rightarrow 0$ the barriers' 
profile become strongly localized  impurities. Furthermore, in 
the case where $V_i=b_i/(\sqrt{2\pi} \sigma_i)$ (for fixed $b_i$), 
the potential~(\ref{Vg}) features $\delta$-like peaks
and can be approximated as:
\begin{equation}
V(x)= \sum_{i=1}^2 b_i \delta(x-l_i).
\label{Vdelta}
\end{equation}

In order to find an effective particle-like equation of motion for the soliton center 
we follow the analysis of Ref.~\cite{gt} (see also Ref.~\cite{sch} for an application 
in the case of Gaussian barriers). In particular, we first seek stationary solutions 
of Eq.~(\ref{nls}), of the form $u=u_b(x){\rm e}^{-i \mu t}$, where the
real function 
$u_b(x)$ represents the spatial profile of the background field, and $\mu
$ is 
the chemical potential. 
Then, it is readily found that $u_b(x)$ satisfies the equation:
\begin{eqnarray}
u_b+\frac{1}{2}\frac{d^2 u_b}{dx^2} -u_b^3 = V(x) u_b, 
\label{cl}
\end{eqnarray}  
where, without loss of generality, we have fixed the chemical potential at $\mu=1$. 

Let us now assume that the barriers' amplitudes are sufficiently small. 
In such a case, when the amplitude $\max|u_b(x)|$ is small, the nonlinear term in Eq.~(\ref{cl}) 
can be neglected and, taking into regard that in the homogeneous case ($V(x)=0$)  
the background amplitude is equal to $1$ when $\mu=1$, 
we look for a solution of Eq.~(\ref{cl}) in the form: 
\begin{eqnarray}
u_b(x)= 1+f(x),
\label{u_b}
\end{eqnarray}
where $f(x)$ incorporates the perturbation by the two Gaussian barriers and has the approximate form: 
\begin{align}
f(x) &= \sqrt{\frac{\pi}{8}} \sum_{i=1}^2 V_i \sigma_i e^{2
\sigma_i^2}   
\nonumber \\
&\times \Bigg \lbrace\Bigg[-1+\rm{erf}\left( \frac{\sigma_i}{\sqrt{2}}\left(2
+\frac{x-l_i}{\sigma_i^2}\right)\right) \Bigg]  
e^{2
\left(x-l_i\right)}    
\nonumber \\
&+\Bigg[-1-\rm{erf}\left(\frac{\sigma_i}{\sqrt{2}}\left(-2
+\frac{x-l_i}{\sigma_i^2}\right)\right) \Bigg] 
e^{-2
\left(x-l_i\right]} \Bigg\rbrace. 
\label{fg}
\raisetag{-.35em}
\end{align}
This way, Eqs.~(\ref{u_b})-(\ref{fg}) describe the spatial profile of  
the effective ground state 
wavefunction, as 
modified by the two barriers. Note that in the limiting case of delta-like 
impurities, the function $f(x)$ can be well
approximated by (see Ref.~\cite{gt} for details):
\begin{eqnarray}
f(x)= -\frac{1}{2} \sum_{i=1}^2 b_i e^{-2
|x-l_i| }.
\label{f_delta}
\end{eqnarray}

To describe the dynamics of a dark soliton on top of this inhomogeneous  
background, we seek  a solution of Eq.~(\ref{nls}) in the form 
\begin{eqnarray}
u(x,t) &=& u_b(x)\exp(-i
t)\upsilon(x,t), 
\label{ud}
\end{eqnarray}
where 
the unknown complex field $\upsilon(x,t)$ represents a dark soliton. Notice that 
in the homogeneous case, the dark soliton wavefunction is given by:
\begin{eqnarray}
\upsilon(x,t) &=& \cos\phi \tanh X +i\sin\phi,
\label{dark}
\end{eqnarray}
where $X \equiv \cos\phi[x - x_0(t)]$ is the soliton coordinate, $\phi$ is the soliton phase angle
$(|\phi|<\pi/2)$ describing the ``darkness'' of the soliton, with 
$\cos\phi$ being the soliton depth ($\phi=0$ and $\phi \ne 0$ correspond to
black and gray solitons, respectively), while $x_0(t)$ and $dx_0/dt = \sin\phi$ denote the position of the soliton center and velocity, respectively.

\begin{figure}[tbp]
\centering
\includegraphics[scale=0.4]{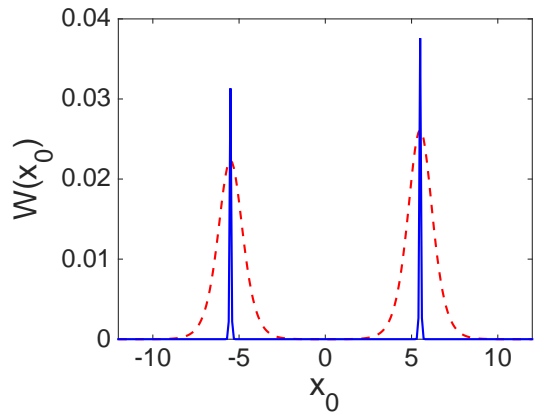} \\
\caption{(Color online) The effective potential $W(x_0)$, in the case of  
two asymmetric Gaussian barriers [dashed (red) line], for $V_1=0.1$, $V_2=0.12$, and 
$\sigma_i=0.5$ ($i=1,2$). Shown also is the limiting case of $\sigma_i\rightarrow 0$, 
corresponding to the Dirac $\delta$ functions [solid (blue) line], for 
$b_i=V_i \sigma_i \sqrt{2 \pi}$, i.e., $b_1=0.12$ and $b_2=0.15$. 
In both cases $l_1=-5.5$ and $l_2=5.5$. }
\label{deltafigs}
\end{figure}

Substituting Eq.~(\ref{ud}) into Eq.~(\ref{nls}),  
a perturbed nonlinear Schr{\"o}dinger equation for the dark soliton wavefunction is obtained. 
To treat analytically the soliton motion, we employ the adiabatic perturbation theory 
developed in Ref.~\cite{yskyang} -- see Appendix A for more details, as well as 
the reviews of Refs.~\cite{book_fr,nl_fr} for applications of this approach in BECs. This way,   
the following equation of motion for the soliton center is obtained \cite{sch}:

\begin{eqnarray}
&&\frac{d^2x_0}{dt^2}=-\frac{dW}{dx_0} \nonumber \\
&&=\sqrt{\frac{\pi}{8}} 
\sum_{i=1}^2 \Bigg\lbrace V_i\sigma_i
{\rm e}^{2
\sigma_i^2}
\int_{-\infty}^{+\infty} \Big[F_{1i}(x)+F_{2i}(x)\Big] dx \Bigg \rbrace, 
\nonumber \\ 
\label{exkin}
\end{eqnarray}
where the functions $F_{1i}(x)$ and $F_{2i}(x)$ are given by:
\begin{subequations}
\begin{eqnarray}
F_{1i} &=&  \left\{-1-\rm{erf} \left[ \frac{\sigma_i}{\sqrt{2}} \left(-2
+\frac{x-l_i}{\sigma_i^2} \right)\right]\right\} 
{\rm e} ^{-2 \left(x-l_i \right)}  
\nonumber \\
&\times&  \Big[ \tanh\left(x-l_i-x_0 \right)-1 \Big]  
\nonumber \\
&\times& \sech^4 \left(x-l_i-x_0\right), 
\label{f1i} \\
F_{2i} &=&  \left\{-1+\rm{erf} \left[ \frac{\sigma_i}{\sqrt{2}}\left( 2
+\frac{x-l_i}{\sigma_i^2} \right)\right]\right\} 
 {\rm e}^{2
 \left(x-l_i \right)} 
\nonumber \\
&\times& \Big[ \tanh\left(x-l_i-x_0\right)+1\Big] 
\nonumber \\
&\times& \sech^4 \left(x-l_i -x_0\right). 
\label{f2i}
\end{eqnarray}
\end{subequations}
The effective potential $W(x_0)$ can then be determined by numerically integrating 
Eq.~(\ref{exkin}). A similar equation of motion for the soliton center can also be 
found in the limiting case of $\delta$-like impurities, namely:
\begin{eqnarray}
&&
\!\!\!\!\!\!
\frac{d^2x_0}{dt^2}=-\frac{dW}{dx_0} 
\nonumber \\
&&
\!\!\!\!\!\!
= \frac{3}{8} \sum_{i=1}^2  \Bigg \lbrace b_i  
\int _{-\infty}^{+\infty} \sgn(x-l_i) 
\sech ^4(x-x_0) {\rm e}^{-2 |x-l_i|} dx  \Bigg \rbrace. 
\nonumber \\
\label{df}
\end{eqnarray}
In this case too, the effective potential $W(x_0)$ can be found upon integrating Eq.~(\ref{df}). 
Nevertheless, $W(x_0)$ can be well approximated by a sum of ${\rm sech}^2$ functions 
(cf. Appendix A, as well as Refs.~\cite{gt,vas} for details) and, thus, $W(x_0)$ reads:
\begin{eqnarray}
W(x_0) \approx \frac{1}{4} \sum_{i=1}^2 b_i {\rm sech}^2 \left(x_0-l_i \right). 
\label{pdelta}
\end{eqnarray}
An example of the shape of the effective potential $W(x_0)$ for both the 
Gaussian [dashed (red) line] and $\delta$-like barriers [solid (blue) line] 
is shown in Fig.~\ref{deltafigs}.

According to the above particle picture for the soliton, 
a dark soliton incident 
towards the first barrier can either 
be reflected or transmitted: if the soliton has a velocity 
$v=dx_0/dt$ and, thus, a kinetic energy:
\begin{equation}
  K = \frac{1}{2} v^2=\frac{1}{2}\sin^2\phi
\label{Ep}
\end{equation}
smaller (greater) than the maximum $W_{\rm max}$ of the effective potential, 
then it will be reflected (transmitted).
For low speeds/kinetic energies, one can further use
the approximation $\sin \phi \approx \phi$ in identifying the relevant
critical point; 
this consideration leads to
$\phi < \phi_{\rm c}$ or $\phi > \phi_{\rm c}$ for reflection or transmission, where
the critical value $\phi_{\rm c}$ of the soliton phase angle is given by
(in the small angle approximation):
\begin{eqnarray}
\phi_{\rm c} &\approx& \sqrt{2 W_{\rm max}}.
\label{kc}
\end{eqnarray}
%

\section{Numerical Results}

\subsection{Bright soliton--two barrier scattering: trapping events}

Before we embark on the case of dark solitons in numerical detail,
we provide a case example of the trapping scenario
that can arise in the case of bright solitons. 
This is intended to partially motivate our corresponding dark soliton results, 
and is also partly shown
because such a scenario was not explicitly illustrated
in Ref.~\cite{Al}.
In particular, we examine Eq.~(\ref{nls}) 
with $s=-1$, in the presence of a Rosen-Morse 
external potential, namely:
\begin{eqnarray}
V(x)=\sum_{i=1}^2  \Big\lbrace -U_i {\rm sech}^2 {\Big[-\alpha_i \left(x-l_i \right)\Big]} \Big\rbrace.
\label{Vb}
\end{eqnarray}
%
Here, $U_i$, $\alpha_i$ and $l_i$ determine the depth, inverse width, and 
the position of the center of the first and second potential, respectively. 
The initial condition which is used has the form of the exact 
bright soliton solution of the homogeneous version of Eq.~(\ref{nls}) for 
$s=-1$; this solution reads:
\begin{eqnarray}
u(x,t=0) &=&  A {\rm e} ^{i v x} \sech \Big[A\left(x-x_0  \right) \Big],
\label{ub} 
\end{eqnarray}
where $A$ sets the amplitude and 
inverse width of the soliton; this parameter  
is taken to be $A=1$. Finally, $v$ and $x_0$ denote, respectively,
the speed and initial position of the soliton. 

\begin{figure}[tbp]
\centering
\includegraphics[scale=0.35]{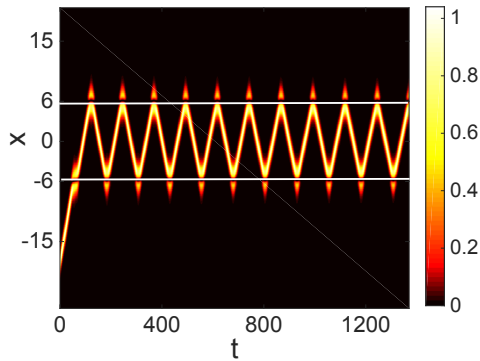} 
\caption{(Color online) Contour plot for 
the evolution of a bright soliton density 
depicting the effect of soliton trapping. 
The soliton's initial position and velocity are 
$x_0=-20$ 
and $v=0.338$, respectively; other parameter values:
$U_1=U_2=3$, $\alpha_1=\alpha_2=1.73$, 
$l_1=-6$, and $l_2=6$. 
}
\label{sc4}
\end{figure}

In Fig.~\ref{sc4}, we develop an 
example, similar to that 
presented in Ref.~\cite{Al}, 
using the following 
parameter values: $U_1=U_2=U=3$, $\alpha_i=1.73$ for $i=1,2$, $l_1=-6$, 
$l_2=6$. For this set of parameters, and for a bright 
soliton incident from left to right, the velocity needed for transmission 
through the first barrier is $v_{cr}=0.336$ 
(in fairly good agreement with the semi-analytical prediction  
$v_{cr}\approx 0.42 U^{-0.18} \approx 0.34$ of Ref.~\cite{Al}). 
The bright soliton is initially located at $x_0=-20$ and has an initial 
velocity slightly larger than the critical, $v=0.338$ in the figure. 
As a result, after it passes the first barrier, the 
soliton loses part of its energy; 
hence, its velocity is reduced to $v^{*}=0.335$, which is 
smaller than the critical value for these (identical) barriers.
As a consequence, 
the soliton falls into a (nearly) periodic state --losing a minimal
amount of its energy after each collision--, remaining
trapped between the two defects. 
Therefore, in addition to the two velocity 
regimes of reflection and transmission, there exists also a third
narrow parametric regime, whereby the radiation from the first
transmission leads to a subcritical incidence speed with respect
to the second barrier which, in turn, allows the soliton to become
indefinitely trapped. 

\subsection{Symmetric potentials}

\begin{figure}[tbp]
\centering
\includegraphics[scale=0.35]{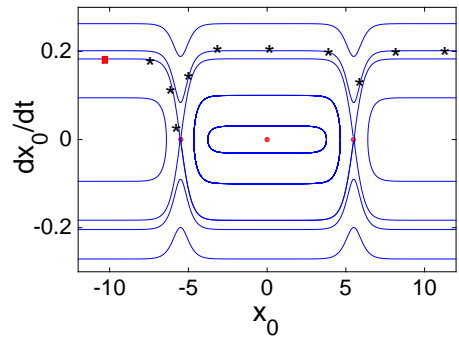}  \\
\includegraphics[scale=0.35]{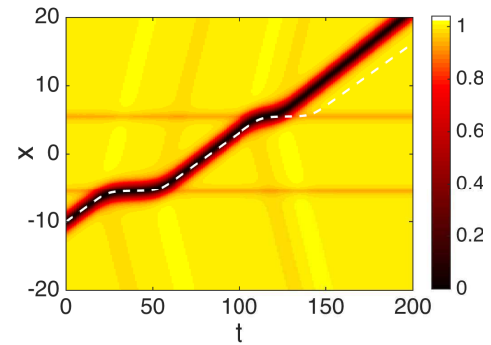} 
\caption{(Color online) The case 
of two symmetric Gaussian barriers, with 
$V_i=0.3$, $\sigma_i=0.1$ ($i=1,2$), 
$l_1=-5.5$ and $l_2=5.5$.  
Top panel: phase plane, where stars depict 
PDE results showing how the dark soliton accelerates upon incidence. 
As a result, while starting essentially on the
stable manifold of the first saddle point, upon collision with
the first barrier accelerates, becoming supercritical with respect
to the second saddle point. 
Bottom panel: contour plot showing the evolution of the dark soliton density 
for the initial condition depicted in the middle panel with the 
(red) square, i.e., $x_0=-10$ and $\phi=0.188>\phi_c=0.183$.
The dashed (white) curve depicts the ODE result,  
which underestimates the distance traveled by 
the dark soliton 
(see relevant discussion in the text).}
\label{sc1}
\end{figure}

Turning now to the case of dark solitons, our first scenario 
refers to 
the case of a symmetric potential, with two Gaussian barriers
bearing the same amplitudes (similarly to the bright soliton case studied above). 
The relevant situation, 
shown in Fig.~\ref{sc1}, 
corresponds to parameter values  
$V_1=V_2=0.3$, $\sigma_1=\sigma_2=0.1$,  
$l_1=-5.5$ and $l_2=5.5$. 
Here, the energy threshold
needed to be overcome by the dark soliton's kinetic energy 
in order for the soliton to be transmitted is 
$W_{\rm max}=0.017$; 
hence, according to Eq.~(\ref{kc}), the critical phase angle for transmission/reflection 
is $\phi_c\approx 0.184$. 

We consider a soliton with initial position 
$x_0=-10$ and phase angle slightly larger than the critical one, i.e. $\phi=0.188$ 
[cf. (red) square point in the phase plane shown in the top panel of Fig.~\ref{sc1}]. 
After the soliton passes the first barrier, 
it emits radiation in the form of sound waves; 
as a result (and this is the fundamental difference in the case
of the dark solitons, which operate as ``negative mass'' particles -- cf., e.g., Refs.~\cite{nl_fr}),
the soliton becomes shallower and thus faster. 
In the bottom panel of Fig.~\ref{sc1}, 
where a contour plot depicting the evolution of the soliton density is shown, 
it is clear that the analytical 
approximation based on the ordinary differential equation (ODE) Eq.~(\ref{exkin}), 
underestimates the soliton velocity: 
the latter, upon
incidence to the second barrier, has increased to $\phi \approx 0.2$.
As a result, the dark soliton not only avoids trapping (entirely
contrary to the case of the bright solitons), but it emerges
with a higher kinetic energy (and hence at a larger distance
than its theoretically predicted counterpart) after the second barrier.
The stars in the phase plane show the corresponding partial differential equation (PDE) 
results [obtained by direct integration of Eq.~(\ref{nls})].
It is clear from the phase plane evolution that while starting
very proximally to the stable manifold of the saddle point at
the first barrier, upon incidence, the soliton accelerates 
(its radiation emission enhances its speed) and, hence, it finds
itself definitively over the heteroclinic orbit associated
with the second barrier, thus escaping to positive infinity.
Qualitatively similar results can also be obtained for 
a variety of different parameters and are representative of the nature
of the dark soliton dynamics. 

\subsection{Asymmetric potentials}

Motivated by the above observations, 
the second scenario that we consider concerns the case of slightly asymmetric defects. 
This allows for regimes where the first defect transmits yet 
the second one reflects. We thus make the choice of the second barrier 
being a little higher than the first one. We observe that, as before, 
the dark soliton becomes shallower after it passes the first 
barrier and it is accelerated  following a different trajectory in
the phase plane. 
Yet, in this case, it may still remain within the region leading 
to reflection from the second barrier.
Nevertheless, when returning to the first, shallower barrier, the
dark soliton is supercritical with respect to it, leading
to its escape towards minus infinity.

\begin{figure}[tp]
\centering
\includegraphics[scale=0.35]{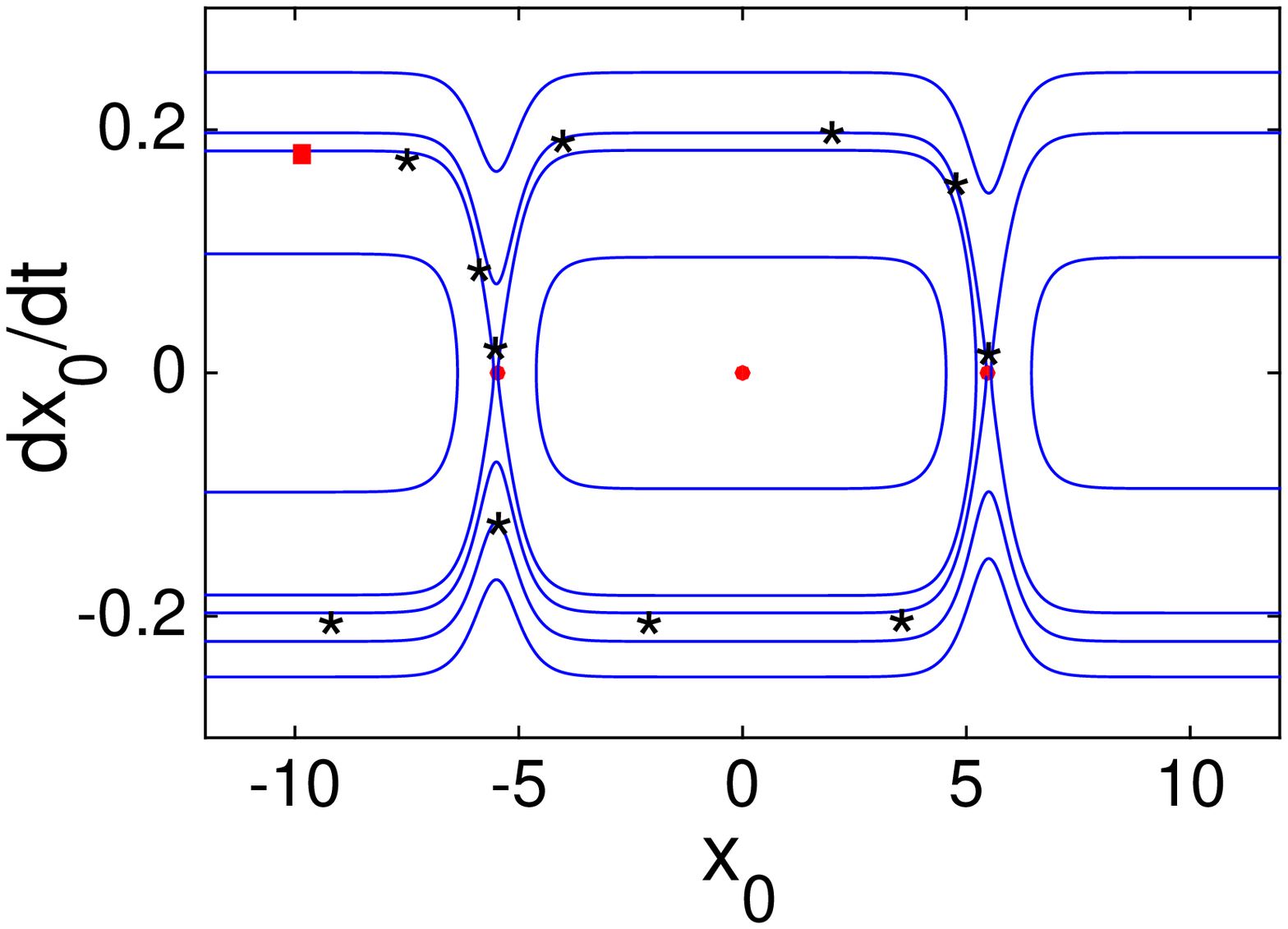} \\
\includegraphics[scale=0.35]{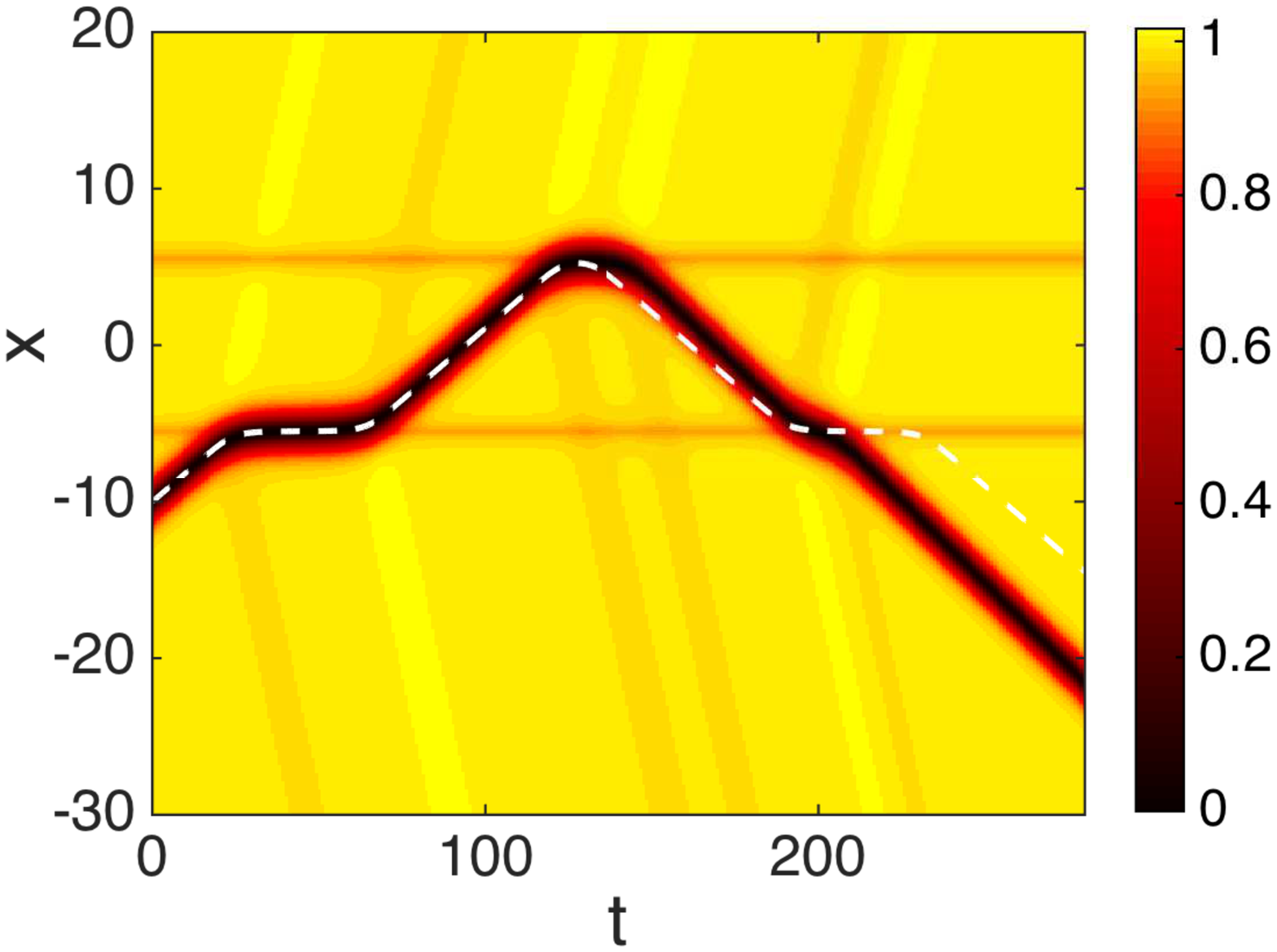}
\caption{(Color online) Similar to Fig.~\ref{sc1}, but for 
two asymmetric Gaussian barriers, with 
$V_1=0.3$, $V_2=0.34$, $\sigma_1=\sigma_2=0.1$,  
$l_1=-5.5$ and $l_2=5.5$.  
}
\label{sc2}
\end{figure}

The above configuration is shown in Fig.~\ref{sc2}, where we have used 
$V_1=0.3$, $V_2=0.34$, $\sigma_1=\sigma_2=0.1$,  
$l_1=-5.5$ and $l_2=5.5$; 
in this case, the energy threshold $W_{\rm max}$ 
is the same as before, i.e., $W_{\rm max}=0.017$.  
We consider a soliton with initial position 
$x_0=-10$ and, as before, a phase angle slightly larger than the critical one, 
i.e., $\phi=0.188$ (cf. square (red) point in the phase plane shown in the 
top panel of Fig.~\ref{sc2}). 
The stars in the phase plane show the corresponding PDE results.   
Indeed, the soliton velocity after it interacts with the first defect
becomes $\phi \approx 0.2$, yet it is still reflected from the
asymmetric (taller) second barrier.

Upon return to the first barrier, the soliton exits the trapping region,
once again with a larger speed than predicted from the particle picture
(dashed line in the figure).
Here, too, qualitatively similar results 
are found for a variety of different parameter values, confirming
the acceleration of the dark soliton as a result of its radiation
emission.

\subsection{Time-dependent impurities}

\begin{figure}[tbp]
\centering
\includegraphics[scale=0.35]{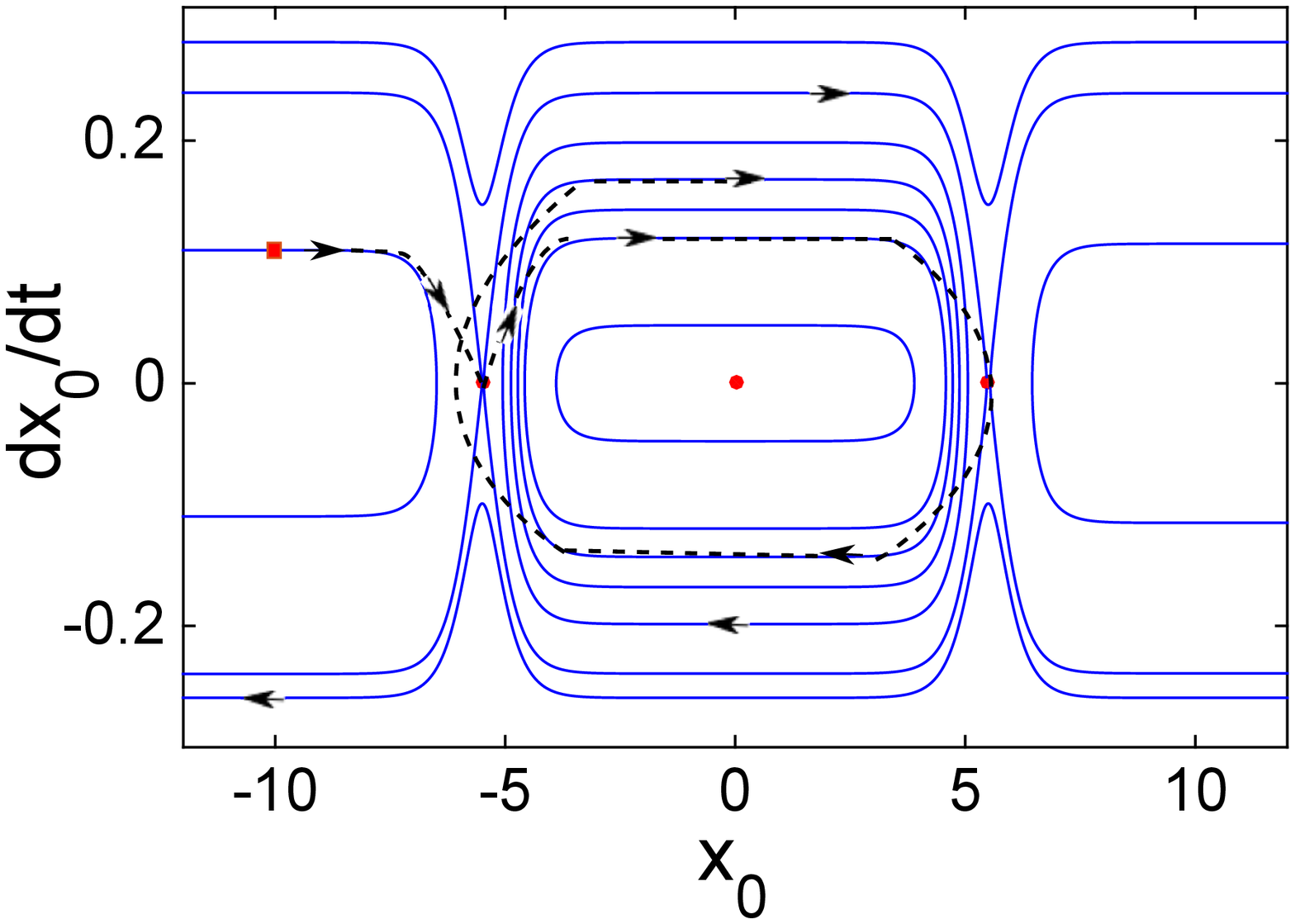} \\
\includegraphics[scale=0.36]{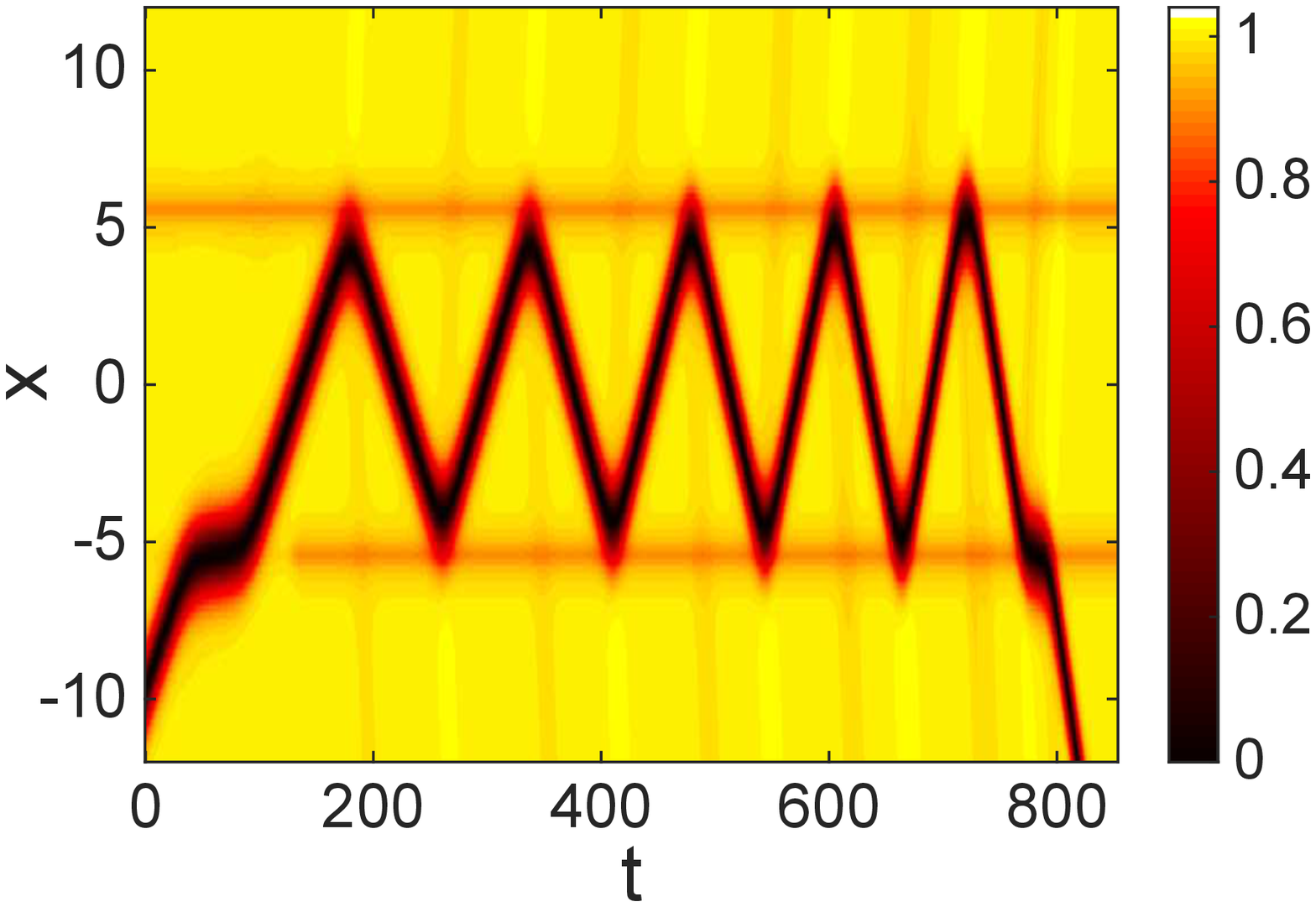} 
\caption{(Color online) 
The case where the first Gaussian impurity is 
time-dependent, and remains repulsive, for   
$V_1=0.1$, $V_1^{*}=0.5$, $V_2=0.5$, 
$\sigma_1=\sigma_2=0.1$,  
$l_1=-5.5$ and $l_2=5.5$.  
Top panel: corresponding phase plane associated with the 
final value of the $V_1$, i.e. $V_1^{*}=0.5$. Dashed (black) 
line and (black) arrows depict respective PDE results showing the 
continuous acceleration of the dark soliton that eventually 
lead to its departure from the trapping region. 
Bottom panel: contour plot showing the evolution of the dark soliton density 
for the initial condition depicted in the top panel with the 
(red) square, i.e., $x_0=-10$, $\phi=0.11$. 
}
\label{sc3}
\end{figure}

As we have already shown in the case examples of Figs.~\ref{sc1} and 
\ref{sc2}, when the soliton interacts with the first defect with a speed close to 
the critical threshold, it passes the first defect, and then 
it becomes shallower (and thus faster). If its velocity is still smaller 
than the critical velocity needed to overcome the 
second barrier it will there be reflected, 
but inevitably it will be finally transmitted through the first barrier. 
Thus, it is not possible to identify 
a regime for dark soliton trapping, similar to the one 
found for bright solitons in the case of 
slightly asymmetric (or symmetric) fixed defects.

Hence, we now focus on the case where the impurities are
--and in particular the first one (from the left) is-- time-dependent.  
In such a case, the external potential is of the following form: 
\begin{equation}
\begin{aligned}
V(x,t) &= V_1(t) \exp{\left[-\frac{1}{2}\left(\frac{x-l_1}{\sigma_1 }\right)^2\right]}  \\
&+ V_2 \exp{\left[-\frac{1}{2}\left(\frac{x-l_2}{\sigma_2 }\right)^2\right]},   \\
V_1(t) &= \frac{1}{2} \Bigg[ \left(V_1^{*}+V_1 \right) +
\left(V_1^{*}-V_1\right) \tanh\Big(\frac{t-t^{*}}{w^{*}} \Big) \Bigg],  
\end{aligned}
\label{V_t}
\end{equation}
where $V_1$ is the (asymptotic) amplitude of the first barrier before the time
$t^*$, while 
$V_1^{*}$ is the (asymptotic) amplitude of the first  Gaussian well after
that time.
Here, we pick $t^{*}$ as a time after the dark soliton is transmitted through the first 
barrier but well before it returns to it.
In turn, $w^{*}$ denotes the (chosen to be short) time interval over which
the first barrier changes value between its asymptotic limits.

We focus on two different cases. In the first case, the left Gaussian impurity
remains repulsive over time, and in the second one it changes sign turning from attractive to repulsive. 

For the first case, cf. Fig.~\ref{sc3}, we have chosen  
$V_1=0.1$, $V_1^{*}=0.5$, $V_2=0.5$, $\sigma_1=\sigma_2=0.1$,  
$l_1=-5.5$ and $l_2=5.5$. 
We consider a soliton with initial position 
$x_0=-10$ and phase angle $\phi=0.11$, 
slightly larger than the critical one. 
After the soliton is transmitted, as expected, from the 
first barrier it is accelerated but it still has a velocity smaller than the critical 
threshold of the second barrier.
Meanwhile, the amplitude of the first (left) Gaussian 
is tuned to become equal to the second (right) Gaussian's amplitude 
(notice that, in experimental realizations
of such barriers~\cite{dic,exp_eng}, this amounts to a straightforward
tuning of the laser beam). 
Hence, when the dark soliton interacts 
again with the first barrier it is reflected from it, 
resulting in an oscillation between the two 
potential barriers. Nevertheless,
as the soliton starts oscillating back and forth now,
being temporarily trapped between the two barriers, 
these oscillations acquire, in fact, progressively smaller period (the soliton
keeps getting faster and faster) due to the emitted radiation; 
as a result, eventually the soliton is 
transmitted through one of the two barriers. 

As observed in Fig.~\ref{sc3}, the soliton completes four oscillations before it escapes. 
We have also tried different parameter values and, as conclusion, 
for smaller amplitude differences, i.e., $V_1^{*}-V_1$, 
we expect less oscillations, while for bigger amplitude differences 
we observe more oscillations. Nevertheless, in a fashion fundamentally 
different than its bright soliton counterpart, the ``anti-particle''
nature of the dark soliton and its progressive acceleration will
always lead to its expulsion from the trapping region.

\begin{figure}[tbp]
\centering
\includegraphics[scale=0.35]{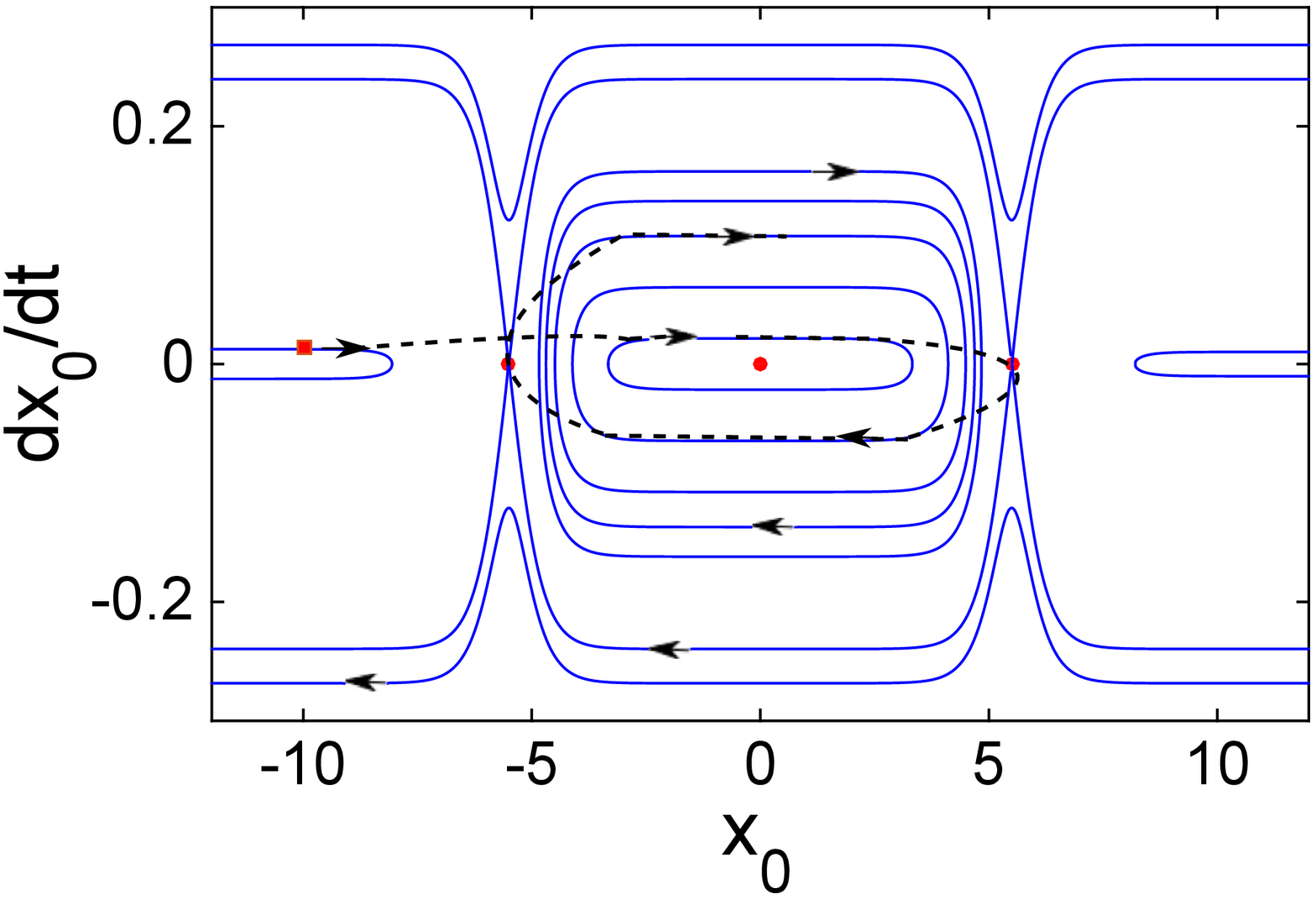} \\
\includegraphics[scale=0.35]{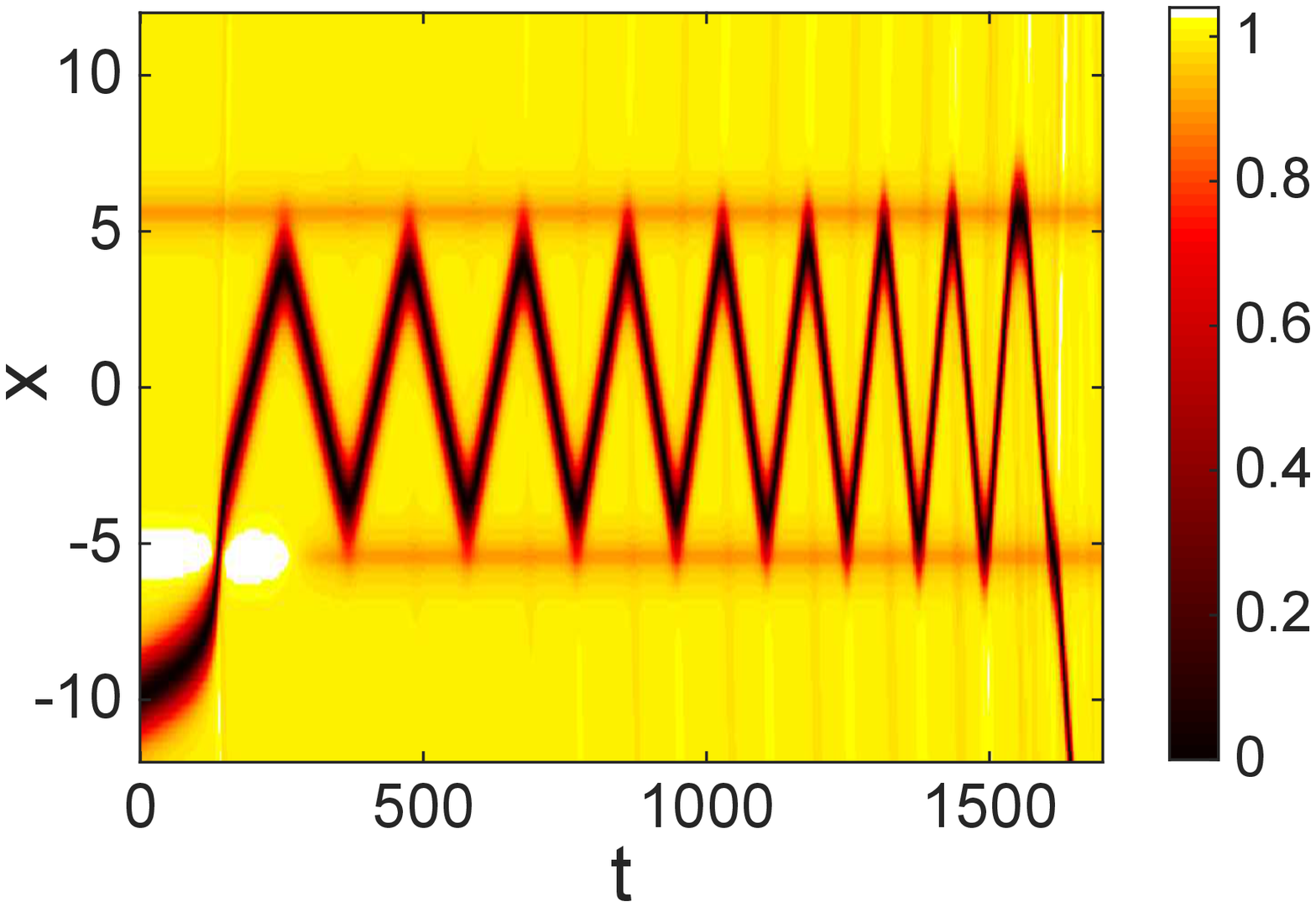} 
\caption{(Color online) Similar to Fig.~\ref{sc3}, but for 
the case where the first Gaussian impurity is time-dependent, while changing its sign, 
for $V_1=-0.5$, $V_1^{*}=0.5$, $V_2=0.5$, 
$\sigma_1=\sigma_2=0.1$,  
$l_1=-5.5$ and $l_2=5.5$. Parameter values for the dark soliton: 
$x_0=-10$, $\phi=0.01$.   
}
\label{sc5}
\end{figure}

This can be also observed in the phase plane 
associated with the final value of the $V_1$, i.e. $V_1^{*}=0.5$, 
shown in the 
top panel of Fig.~\ref{sc3}. 
The dashed (black) line depicts the 
first oscillations of the corresponding PDE results. 
After the soliton interacts with each barrier, it is getting faster 
following different orbits of an increasing speed
--and amplitude-- in the phase plane (see (black) arrows).   
This leads the soliton to follow a spiral trajectory
that eventually will allow its departure from the 
region between the two barriers. 

Let us now consider the second case where, initially, the first defect provides 
a well rather than a barrier, facilitating the dark soliton's transmission.
Relevant results are shown in Fig.~\ref{sc5}, where we have used 
$V_1=-0.5$, $V_1^{*}=0.5$, $V_2=0.5$, $\sigma_1=\sigma_2=0.1$, 
$l_1=-5.5$ and $l_2=5.5$. 
Again, the phase plane associated with the final value of the $V_1$, i.e. $V_1^{*}=0.5$, 
is shown in the 
top panel of the same figure. 
The soliton is initially placed at  
$x_0=-10$ and it has a phase angle much smaller than before, 
equals to $\phi=0.01$. 

Here, 
after the interaction with the first barrier, the soliton's energy 
does not change significantly; thus, the soliton is reflected from the second barrier
and returns to the first defect which, in the meantime, 
has become a barrier as well. Given its minimal speed,
the dark soliton is able to execute 
many oscillations between the two defects in this case, 
until it is finally able to escape due to the same mechanism as
discussed above.

\section{Conclusions and Future Challenges}

In the present work we have revisited the phenomenon of dark soliton
scattering in the presence of a two-defect potential. This was
partially motivated by the intriguing findings for bright solitons 
in the earlier work of Ref.~\cite{Al}:
there, phenomena such as a ``soliton diode'' were identified in the case of asymmetric
barriers, 
and the possibility of soliton trapping was also predicted 
even for symmetric potential barriers.

We developed a particle picture allowing us to explore the motion
of dark solitons within the realm of two barriers. The most
fundamental feature identified was that a scenario of two
equal barriers can never trap a dark soliton, fundamentally
contrary to what is the case with a bright one. This is because
the ``anti-particle'' character of the dark soliton leads to its
increase in speed, upon scattering from the first defect, hence
it will always overcome the second defect. Making the barriers asymmetric
will not help either, because although the second barrier may reflect
the soliton, the first one will always transmit it upon its return, 
so trapping cannot be achieved. The only possibility for the latter 
is a phenomenology employing time-dependent barriers. Even here, 
the dark soliton cannot be trapped indefinitely, yet it can 
be trapped for long 
intervals of evolution time, executing many oscillations.

We conjecture that given a very general potential supported sufficiently 
far from an initial dark soliton, the dark soliton is nonetheless asymptotically 
stable in that the dynamics lead to a nearby dark soliton moving left or right, 
depending upon the initial phase angle $\phi$. 
Recent progress on the Gross-Pitaevskii equation with delta function potentials 
was made in Ref.~\cite{delta_cauchy}, where orbital stability of stationary dark 
solitons in the setting of a single $\delta$ function potential was proved. 
The mathematical theory for this setting initiated with the works of Refs.~\cite{ger1,ger2}.

There are numerous interesting questions for future study that are emerging
from the present work. For one thing, it would be especially interesting
to explore whether the radiation from the dark soliton could in any 
analytical (see, e.g., Refs.~\cite{dep1}) or even numerical (see, e.g., Refs.~\cite{pr1}) 
way be captured. If, especially, analytical results were available then the radiation could be  
incorporated in the equations of motion, so as to enable a
quantitative characterization of the soliton dynamics in the presence 
of sound waves. Additionally, it would be quite relevant to explore
somewhat systematically the phenomenology of scattering in the presence
of a double well, rather than a double barrier (see relevant work in Ref.~\cite{pr2}). 
Finally, extending
the relevant considerations to the realm of a larger number of components
or a larger dimension would be important. In the former, the examination
of the interaction of dark-bright solitons with multiple barriers
would be a natural next step~\cite{RIP}, while in the latter, exploring
the interaction of vortices with such potentials would be of
particular interest \cite{book_fr}.

\appendix
\section{Particle dynamics with $\delta$ function defects}

Here we provide some details on the perturbation theory for dark soliton 
dynamics in the presence of $\delta$ function defects. A similar analysis 
can also be applied for the case of Gaussian barriers, leading to the results 
presented in Sec.~II.

Our starting point will be the perturbed NLS equation stemming from the 
substitution of Eq.~(\ref{ud}) into Eq.~(\ref{nls}): 
\begin{equation}
i \upsilon_t + \frac12 \upsilon_{xx} -(|\upsilon|^2 -1) \upsilon = P(\upsilon),
\label{pertnls}
\end{equation}
where perturbation 
$P(\upsilon)=2f\left(1-|\upsilon |^2  \right)\upsilon - (df/dx)\upsilon_x$ 
(with $f(x)$ given in Eq.~(\ref{f_delta})) can be expressed as follows:
\begin{equation} 
P (\upsilon) = 
\sum_{i=1}^2 \Bigg \lbrace  b_i e^{-2 u_0 |x-l_i|} 
\bigg[ \left(1-|\upsilon |^2  \right)\upsilon - \sgn(x-l_i)
 \upsilon _x  \bigg] \Bigg \rbrace.
\label{Pu}
\end{equation}
We now employ the adiabatic perturbation theory \cite{book_fr,nl_fr,yskyang}, 
according to which, the functional form of the soliton remains unchanged, 
but its parameters $x_0$ and $\phi$ become slowly varying functions of time $t$. 
In other words, we seek solutions of Eq.~(\ref{pertnls}) of the form 
\begin{equation}
\upsilon(x,t) = \cos\phi (t) \tanh X + i \sin \phi(t), 
\label{ups}
\end{equation}
with $X= \cos\phi (t)[x - x_0 (t)]$ and 
\begin{equation}
\label{x0def}
x_0 (t) = \int_0^t \sin\phi(s) ds.
\end{equation}
Then, following energy considerations \cite{book_fr,nl_fr,yskyang}, the following 
equation for the evolution of the soliton phase angle can be derived:
\begin{equation}
\label{phidyn}
\frac{d\phi}{dt} = 
\frac{1}{2 \cos^2 \phi \sin \phi} \text{Re} \ \left( \int_{-\infty}^\infty 
P(\upsilon) \bar{\upsilon}_t dx \right),
\end{equation}
with $P(\upsilon)$ and $\upsilon$ defined, respectively, in Eqs.~(\ref{Pu}) and (\ref{ups}) 
above. Notice 
that only the real part of $\upsilon$ is $x$-dependent, which 
greatly simplifies calculations.
Computing the integral in Eq.~(\ref{phidyn}), and using the assumption that 
the solitons are nearly black (i.e., $\phi \rightarrow 0$), we end up with 
the result:
\begin{equation}
\frac{d\phi}{dt} = \frac{3}{8} \sum_{i=1}^2  \Bigg \lbrace b_i  
\int _{-\infty}^{+\infty} \sgn(x-l_i) 
\sech ^4(x-x_0) {\rm e}^{-2 |x-l_i|} dx  \Bigg \rbrace. 
\label{dphidt}
\end{equation}
The integral in the above equation, which is of the form 
$I=\int_{-\infty}^{+\infty} \left[\sgn(x)\sech^4(x-x_0)e^{-2|x|}\right] dx$,
can be evaluated using that
$I = (2/3)\sech^2(x_0)\tanh(x_0)$.
Combining 
this simplified equation with Eq.~(\ref{x0def}), we obtain -- from 
the equation of motion for the soliton center -- Eq.~(\ref{pdelta}), namely 
the effective potential.

\begin{figure}[tbp]
\centering
\includegraphics[scale=0.35]{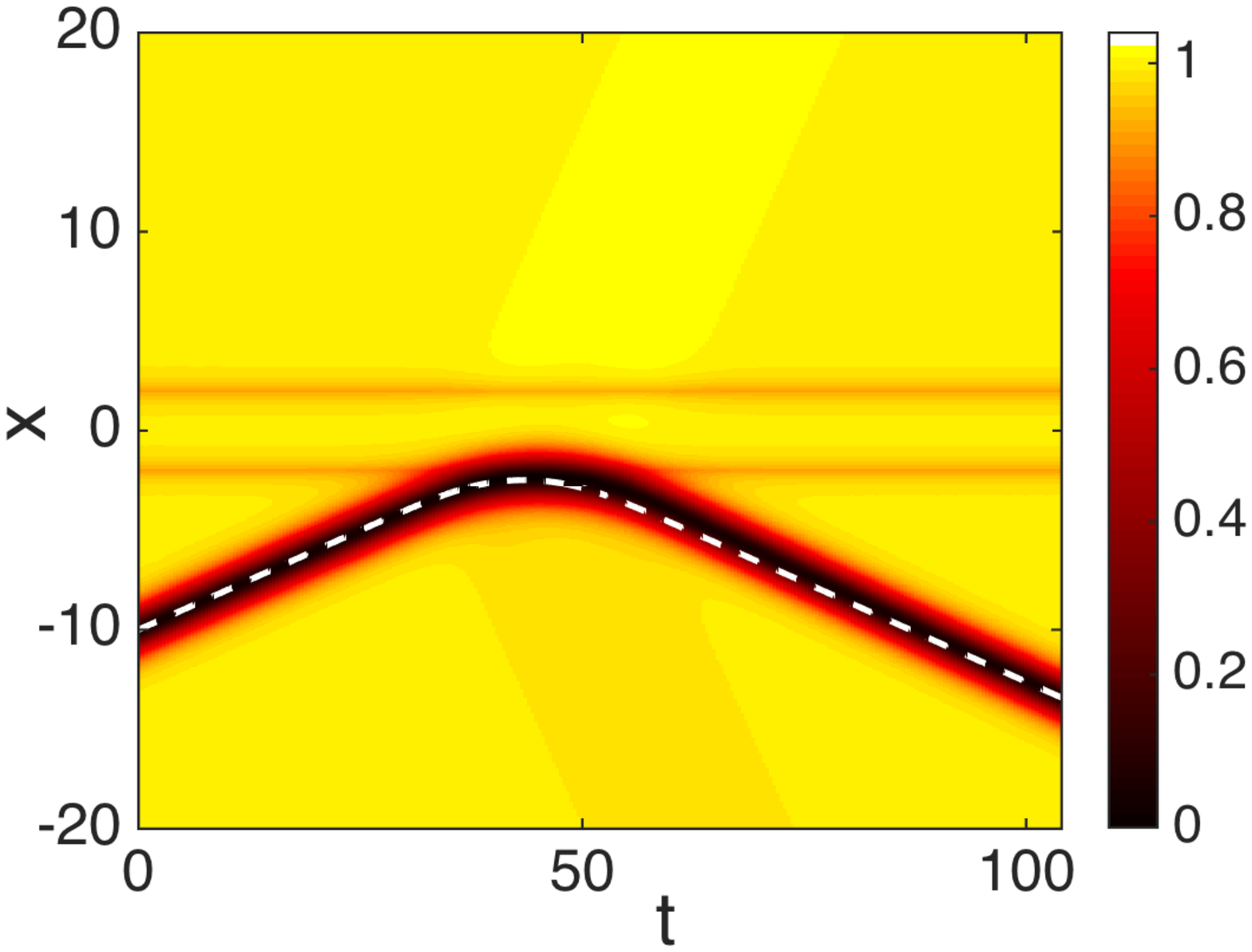} \\
\includegraphics[scale=0.35]{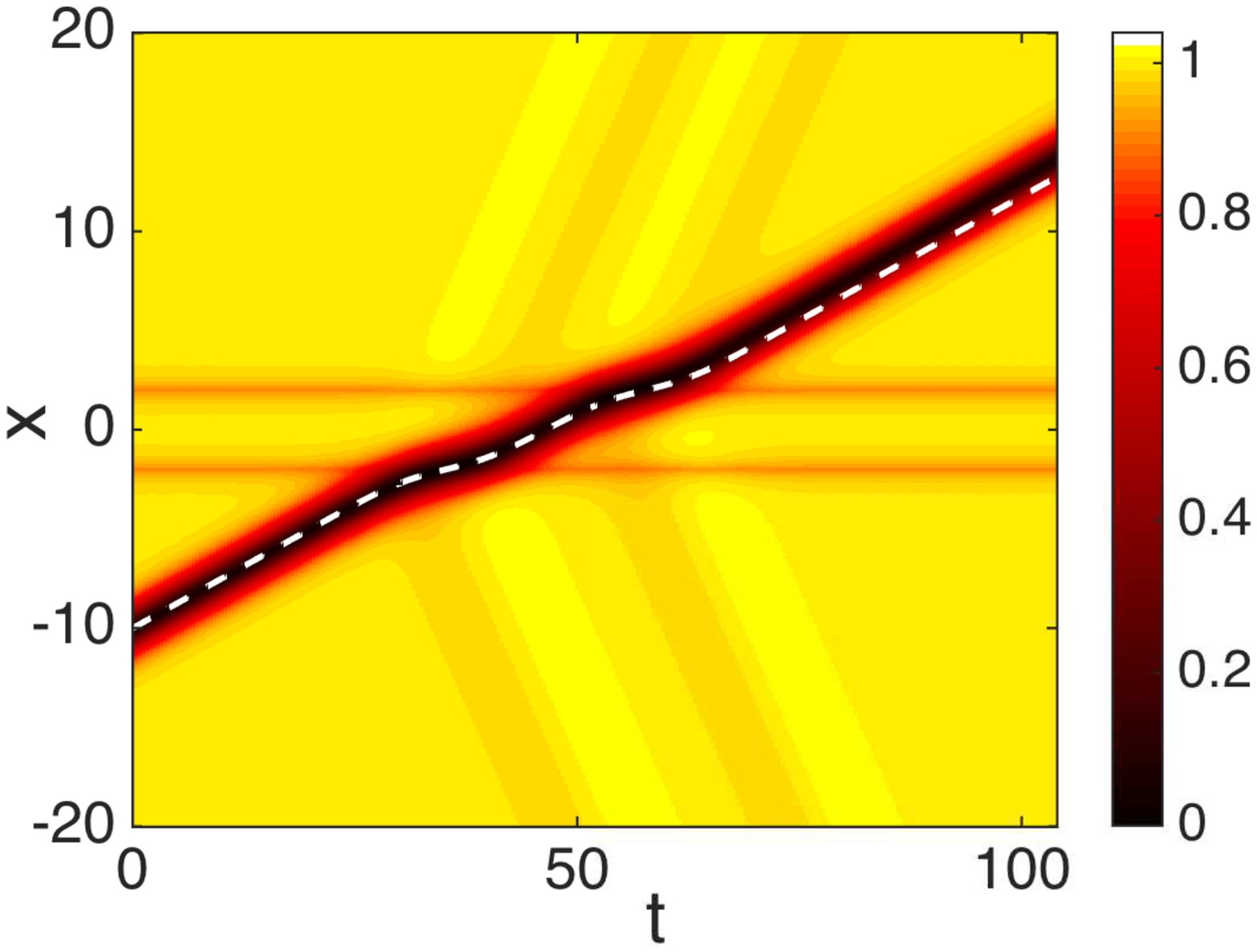} 
\caption{(Color online) 
Contour plots showing the 
evolution of the dark soliton density 
for two symmetric Dirac delta barriers located at $l_1=-2$ and $l_2=2$; 
the soliton is initially placed at  
$x_0=-10$ and has a phase angle $\phi=0.2$ (top panel) and $\phi=0.25$ (bottom panel). 
The dashed lines show the particle picture from Eq.~(\ref{df}). 
}
\label{deltafigs1}
\end{figure}

Finally, it is worth noticing that a similar feature to the Gaussian barriers 
occurs in the case of the $\delta$ potential 
barriers as well. In the case of symmetric potentials with $b_1=b_2=b$ for instance, 
it is found that,  
for a given value of $b$, there is a critical value of $\phi$ 
such that below this critical value the dark soliton 
is reflected at the first barrier,
and above the dark soliton is transmitted through both barriers.
A typical example 
is shown in Fig.~\ref{deltafigs1}; 
note that 
the full $\delta$ potential is simulated using a finite element decomposition, similar to that proposed in Ref.~\cite{holmer2}, which allows the $\delta$ functions to enter in a non--approximate
form through the weak formulation of equation \eqref{nls}.

%


\end{document}